# Disentanglement dynamics of interacting two qubits and two qutrits in an *XY* spin-chain environment with the Dzyaloshinsky-Moriya interaction

Ming-Liang Hu[*]

*School of Science, Xi'an University of Posts and Telecommunications, Xi'an 710061, China*

**Abstract:** We investigate disentanglement dynamics of two coupled qubits and qutrits which interact uniformly to a general *XY* spin-chain environment with the Dzyaloshinsky–Moriya (DM) interaction. We obtained exact expression of the time evolution operator and analyzed the dynamical process of the decoherence factors. Through explicitly calculating the concurrence and the negativity, we examined disentanglement behaviors of two coupled central spins evolve from different initial pure states, which are found to be nontrivially different from those of the uncorrelated ones, in particular, the enhanced decay of the entanglement induced by quantum criticality of the surrounding environment may be broken by introducing self-Hamiltonian of the central spin system. Moreover, the DM interaction may have different influences on decay of the entanglement depending on the strength of the system-environment coupling, the anisotropy of the environmental spin chain and the intensity of the transverse magnetic field, as well as the explicit form of the initial states of the central spin system.

## 1. Introduction

Entanglement is one of the most essential features of quantum mechanics, and it also plays an important role in quantum information processing (QIP). However, entanglement of quantum state is very fragile because real quantum systems are inevitably coupled to the environmental degrees of freedom, and these unavoidable couplings often result in the dissipative evolution of quantum coherence and loss of useful entanglement when particles propagate or the computation evolves [1–3]. The decoherence process constitutes indeed one of the major obstacles baffling the practical construction of a quantum computer. However, for practical quantum computation, the devices are inevitably subject to decoherence and decay processes, no matter how much they may be screened from the surrounding environment. Thus, a deep understanding of the disentanglement process of the system of interest is of both theoretical and experimental significance.

Since the coupling between a central quantum system and its surrounding environment leads to decoherence of the system, which may tamper with or even wreck the QIP tasks, it is nature for us to consider the dynamical process of degradation of entanglement due to decoherence. As for decoherence caused by different spin-chain environments, much progress has been achieved so far [4–19], and these investigations can be separated into at least two kinds. The first kind is mainly about investigating effects of the uncorrelated environment (i.e., constituents of the environment have no intra-interactions) on system dynamics [4–6], while the second kind is about investigating effects of the correlated environment on quantum dynamics of the system [7–19], for in the real situation, particles constitute the environment may have interactions with each other.

---
[*] *E-mail*: mingliang0301@xupt.edu.cn (M.-L. Hu)

Decoherence induced by the uncorrelated spin environment has been discussed by several researchers in the past few years. Cucchietti *et al.* considered problem of decoherence caused by a collection of independent environmental spins [5]. Hamdouni examined entanglement dynamics of two spin-half particles which are embedded in separate spin star environments [6]. Decoherence induced by the correlated spin environment has also been discussed extensively in recent years. Quan *et al.* considered influence of the Ising-type correlated environment on the Loschmidt echo (LE), and found that the quantum critical behavior of the environmental system strongly affects its capability of enhancing the decay of the LE [7]. Since then the LE of a coupled system consisting of a central spin (or two central spins) and its surrounding environment [8–10] has been discussed extensively. The results show that the decaying behavior of the LE is sensitively affected by the anisotropy of the environment. In particular, Rossini *et al.* demonstrated that in the short-time region the LE decays as a Gaussian, while up to the long time limit, it approaches an asymptotic value which strongly depends on the strength of the transverse magnetic field [10]. Disentanglement process induced by coupling of the central system with an Ising-type spin-chain environment has been discussed in Refs. [11–14], in which the authors concentrated on the relationship between quantum criticality of the surrounding environment and decoherence of the central system. Their results revealed evidently that the entanglement decay can be best enhanced by the quantum phase transition (QPT) of the environment. Besides these exciting progresses, decoherence influences induced by the *XY* spin-chain environment [15,16] and the *XY* spin-chain environment with the Dzyaloshinsky–Moriya (DM) interaction [17–19] on disentanglement evolution of the central spin system has also been discussed recently.

In this paper, as an extension of the work [13], we present a theoretical investigation of the disentanglement process of both two qubits and two qutrits coupled uniformly to an *XY* spin-chain environment with the DM interaction. Different from the previous works [11–15,17,18] in which the authors considered central spin system with constituents without interactions with each other (i.e., the central spin system has no self-Hamiltonian), here we concentrate on disentanglement dynamics of the correlated central spins described by a general *XXZ* interaction for neglecting the self-Hamiltonian of the central system may not always be a reasonable approximation [5,20]. For example, Cucchietti *et al.* [5] have shown that the decoherence factor displays a Gaussian decay when there is no self-Hamiltonian for the system, while in the presence of self-Hamiltonian, the decay is predominantly a power law. Our model considered in this work includes both the effects of the evolution of the central system and its coupling to the surrounding spin environment, the results revealed that under this consideration, the disentanglement behavior will be nontrivially different from those of the uncorrelated ones, in particular, the enhanced decay of the entanglement induced by quantum criticality of the surrounding environment may be broken by introducing self-Hamiltonian of the central spin system.

## 2. Solution of the model

The total Hamiltonian of the composite system we considered in this paper is given by

$$\hat{H} = \hat{H}_S + \hat{H}_E + \hat{H}_I, \quad (1)$$

where $\hat{H}_S$ and $\hat{H}_E$ denote the self-Hamiltonian of the central spin system and that of the environmental spin chain, respectively, $\hat{H}_I$ denotes the interaction Hamiltonian between the central spin system and its surrounding environment. The explicit forms of $\hat{H}_S$, $\hat{H}_E$ and $\hat{H}_I$ are given by

$$\hat{H}_S = s_1^x s_2^x + s_1^y s_2^y + \Delta s_1^z s_2^z + h_1 s_1^z + h_2 s_2^z,$$

$$\hat{H}_E = \sum_{l=-M}^{M} \left[ \frac{1+\gamma}{2} \sigma_l^x \sigma_{l+1}^x + \frac{1-\gamma}{2} \sigma_l^y \sigma_{l+1}^y + \lambda \sigma_l^z + \vec{D} \cdot (\vec{\sigma}_l \times \vec{\sigma}_{l+1}) \right], \qquad (2)$$

$$\hat{H}_I = g(s_1^z + s_2^z) \sum_{l=-M}^{M} \sigma_l^z.$$

Here $\Delta$ and $h_i$ ($i=1, 2$) in the first line of Eq. (2) denote the degree of the anisotropy of the central spin system along the $z$-direction and the intensity of the transverse magnetic field applied to the two central spins $s_1$ and $s_2$, which are transversely coupled to all spins in the environment through the interaction $H_I$. $\sigma_l^\alpha$ ($\alpha = x, y, z$) in the second and third lines are the familiar Pauli operators acting on the $l$th spin, with the total number of spins in the environment given by $N = 2M + 1$. The parameters $\lambda$ and $\gamma$ characterize the intensity of the transverse magnetic field applied to the environment and the anisotropy in the in-plane interaction of the environment, and $\vec{D}$ is the DM vector coupling which arises from spin-orbit coupling [21,22]. The parameter $g$ in the interaction Hamiltonian characterizes the coupling strength between the two central spins and their surrounding environment. Interaction Hamiltonian of the form of $\hat{H}_I$ was proposed as a simple model of decoherence historically [23] and have gained additional importance in recent years due to its relevance to QIP [13–18].

If one defines an operator-valued parameter $\hat{\Lambda} = \lambda + g(s_1^z + s_2^z)$, and choose $\vec{D} = D\vec{z}$ (i.e., the DM interaction is imposed along the $z$-direction), then the combined Hamiltonian $\hat{H}_{EI} = \hat{H}_E + \hat{H}_I$ becomes

$$\hat{H}_{EI} = \sum_{l=-M}^{M} \left[ \frac{1+\gamma}{2} \sigma_l^x \sigma_{l+1}^x + \frac{1-\gamma}{2} \sigma_l^y \sigma_{l+1}^y + \hat{\Lambda} \sigma_l^z + D(\sigma_l^x \sigma_{l+1}^y - \sigma_l^y \sigma_{l+1}^x) \right]. \qquad (3)$$

To diagonalize the Hamiltonian $\hat{H}_{EI}$, we follow the standard procedure by introducing the conventional Jordan-Wigner (JW) transformation [24]

$$\sigma_l^x = \prod_{m<l}(1-2c_m^\dagger c_m)(c_l + c_l^\dagger), \ \sigma_l^y = -i\prod_{m<l}(1-2c_m^\dagger c_m)(c_l - c_l^\dagger), \ \sigma_l^z = 1 - 2c_l^\dagger c_l, \qquad (4)$$

which maps spins to one-dimensional spinless fermions with creation and annihilation operators $c_l^\dagger$ and $c_l$. By inserting Eq. (4) into Eq. (3), we obtain the resulting Hamiltonian as

$$\hat{H}_{EI} = \sum_{l=-M}^{M} \left[ (1+2iD)c_l^\dagger c_{l+1} + (1-2iD)c_{l+1}^\dagger c_l + \gamma(c_{l+1} c_l + c_l^\dagger c_{l+1}^\dagger) + \hat{\Lambda}(1-2c_l^\dagger c_l) \right]. \qquad (5)$$

Next we introduce the Fourier transformation of the fermionic operators $c_l$ described by $d_k = (1/\sqrt{N})\sum_l c_l e^{-ilx_k}$, where $k = -M,\ldots,M$ and $x_k = 2\pi k/N$, with $N = 2M+1$ being the number of spins in the environment. The Hamiltonian $\hat{H}_{EI}$ can be diagonalized by transforming the fermionic operators into momentum space and then using the Bogoliubov transformation. The final results is

$$\hat{H}_{EI} = \sum_k \Omega_k (b_k^\dagger b_k - 1/2), \qquad (6)$$

where the energy spectrum $\Omega_k$ is given by

$$\Omega_k = 2[\epsilon_k - 2D\sin(x_k)], \qquad (7)$$

with $\epsilon_k = \sqrt{[\cos(x_k) - \hat{\Lambda}]^2 + \gamma^2 \sin^2(x_k)}$. The corresponding Bogoliubov-transformed fermion operators are defined by

$$b_k = \cos\frac{\theta_k}{2}d_k - i\sin\frac{\theta_k}{2}d_{-k}^\dagger, \tag{8}$$

with angles $\theta_k$ satisfying

$$\theta_k = \arcsin\left[\frac{-\gamma\sin(x_k)}{\epsilon_k}\right]. \tag{9}$$

Here in order to obtain the time evolution operator, we adopt the procedure of Refs. [7,13] by introducing the following pseudospin operators $\sigma_{k\alpha}$ ($\alpha = x, y, z$) as

$$\sigma_{kx} = d_k^\dagger d_{-k}^\dagger + d_{-k}d_k, \quad \sigma_{ky} = -id_k^\dagger d_{-k}^\dagger + id_{-k}d_k, \quad \sigma_{kz} = d_k^\dagger d_k + d_{-k}^\dagger d_{-k} - 1. \tag{10}$$

By combination of Eqs. (6), (8) and (10), one can re-express the Hamiltonian $\hat{H}_{EI}$ as

$$\hat{H}_{EI} = \sum_{k>0} e^{i(\theta_k/2)\sigma_{kx}}(\Omega_k \sigma_{kz})e^{-i(\theta_k/2)\sigma_{kx}} + \left[1 - \hat{\Lambda} - 2D\hat{\kappa}\sin(x_k)\right]\sigma_{0z}, \tag{11}$$

where the operator $\hat{\kappa} = (1-\hat{\Lambda})/|1-\hat{\Lambda}|$.

From the above equation and in units of $\hbar$, the time evolution operator $U_{EI}(t) = e^{-i\hat{H}_{EI}t}$ can be obtained explicitly as

$$U_{EI}(t) = e^{-i[1-\hat{\Lambda}-2D\hat{\kappa}\sin(x_k)]\sigma_{0z}t}\prod_{k>0}e^{i(\theta_k/2)\sigma_{kx}}e^{-it\Omega_k\sigma_{kz}}e^{-i(\theta_k/2)\sigma_{kx}}. \tag{12}$$

Obviously, the expression for $U_{EI}(t)$ based on a general XY spin-chain environment with the DM interaction is analogous to that based on the pure Ising model which has been previously reported [13]. The differences come from two aspects. The first is that there are contributions of the DM interaction in the first exponential term of Eq. (12), and the second is the explicit form of the energy spectrum $\Omega_k$ and the angle $\theta_k$ for the Bogoliubov transformation [see Eqs. (7) and (9)]. Due to the obvious differences between these two environment models, one may expects that the decoherence process of the central spin system in the present case will include new features characteristic of the XY spin chain with DM interaction.

Moreover, it can be shown that $[\hat{H}_S, \hat{H}_{EI}] = 0$, i.e., the Hamiltonian $\hat{H}_S$ and $\hat{H}_{EI}$ commute with each other, thus the time evolution operator $U(t)$ for the combined system environment can be written as $U(t) = U_S(t)U_{EI}(t)$, with $U_S(t) = e^{-i\hat{H}_S t}$. Explicitly knowing the expression of $U_S(t)$, one can investigate disentanglement dynamics of the central spin system.

## 3. Disentanglement dynamics of two qubits

For the case that the central system constitutes of two qubits, $s_1$ and $s_2$ contained in the self-Hamiltonian $\hat{H}_S$ [see Eq. (2)] represent the spin-1/2 operators. Its eigenvalues can be obtained straightforwardly as

$$\varepsilon_{0,3} = \frac{\Delta}{4} \pm \frac{h_1+h_2}{2}, \quad \varepsilon_{1,2} = -\frac{\Delta}{4} \pm \frac{\csc(\theta)}{2}, \tag{13}$$

with the corresponding eigenstates given by

$$|\varphi_0\rangle = |00\rangle, \quad |\varphi_1\rangle = \cos(\theta/2)|01\rangle + \sin(\theta/2)|10\rangle,$$
$$|\varphi_2\rangle = \sin(\theta/2)|01\rangle - \cos(\theta/2)|10\rangle, \quad |\varphi_3\rangle = |11\rangle, \tag{14}$$

where $\theta = \cot^{-1}(h_1 - h_2)$, with $|0\rangle$ and $|1\rangle$ represent the state of spin up and down, respectively.

From the above two equations, one can obtain the time evolution operator $U_S(t)$ as

$$U_S(t) = \sum_l e^{-i\varepsilon_l t} |\varphi_l\rangle\langle\varphi_l|. \tag{15}$$

We consider disentanglement dynamics of the initial two-qubit pure state which is in a general form as

$$|\psi_S(0)\rangle = a_0|00\rangle + a_1|01\rangle + a_2|10\rangle + a_3|11\rangle, \tag{16}$$

where the complex coefficients satisfying the normalization condition $\Sigma_i |a_i|^2 = 1$. Furthermore, we assume the initial state of the environment to be $|\psi_E(0)\rangle = |0\rangle_{k=0} \otimes_{k>0} |0\rangle_k |0\rangle_{-k}$, where $|0\rangle_k$ denotes the vacuum state of the $k$th mode $d_k$ in the momentum space [13], i.e., $d_k |0\rangle_k = 0$ for any operator $d_k$. For simplicity, we consider the initial state of the combined system environment as $|\Psi(0)\rangle = |\psi_S(0)\rangle \otimes |\psi_E(0)\rangle$, i.e., the initial state is in a product form and there is no initial entanglement between the system and the environment. By applying the time evolution operator $U(t)$, we obtain the system-environment state vector at an arbitrary time $t$ as

$$\begin{aligned}|\Psi(t)\rangle = &\, a_0 e^{-i\varepsilon_0 t}|00\rangle \otimes U_{EI}^{(0)}|\psi_E(0)\rangle + a_1(b_1|01\rangle + b_2|10\rangle) \otimes U_{EI}^{(1)}|\psi_E(0)\rangle \\ &+ a_2(b_2|01\rangle + b_3|10\rangle) \otimes U_{EI}^{(2)}|\psi_E(0)\rangle + a_3 e^{-i\varepsilon_3 t}|11\rangle \otimes U_{EI}^{(3)}|\psi_E(0)\rangle,\end{aligned} \tag{17}$$

where the coefficients $b_1 = [e^{-i\varepsilon_1 t}\cos^2(\theta/2) + e^{-i\varepsilon_2 t}\sin^2(\theta/2)]$, $b_2 = \sin(\theta)(e^{-i\varepsilon_1 t} - e^{-i\varepsilon_2 t})/2$ and $b_3 = [e^{-i\varepsilon_1 t}\sin^2(\theta/2) + e^{-i\varepsilon_2 t}\cos^2(\theta/2)]$. The unitary operators $U_{EI}^{(i)}$ ($i=0,1,2,3$) can be obtained from the unitary operator $U_{EI}(t)$ by replacing the operator $\hat{\Lambda}$ with numbers $\Lambda_{0,3} = \lambda \pm g$ and $\Lambda_{1,2} = \lambda$.

Tracing out the degrees of the environment, we obtained the reduced density matrix of the two central spins in the standard basis $\{|00\rangle, |01\rangle, |10\rangle, |11\rangle\}$ as

$$\rho_S(t) = \begin{pmatrix} |a_0|^2 & a_0\mu_1^* F_{01} e^{-i\varepsilon_0 t} & a_0\mu_2^* F_{01} e^{-i\varepsilon_0 t} & a_0 a_3^* F_{03} e^{i\varepsilon_{30} t} \\ a_0^*\mu_1 F_{01}^* e^{i\varepsilon_0 t} & |\mu_1|^2 & \mu_1\mu_2^* & a_3^*\mu_1 F_{13} e^{i\varepsilon_3 t} \\ a_0^*\mu_2 F_{01}^* e^{i\varepsilon_0 t} & \mu_1^*\mu_2 & |\mu_2|^2 & a_3^*\mu_2 F_{13} e^{i\varepsilon_3 t} \\ a_0^* a_3 F_{03}^* e^{i\varepsilon_{03} t} & a_3\mu_1^* F_{13}^* e^{-i\varepsilon_3 t} & a_3\mu_2^* F_{13}^* e^{-i\varepsilon_3 t} & |a_3|^2 \end{pmatrix}, \tag{18}$$

with $\mu_1 = a_1 b_1 + a_2 b_2$, $\mu_2 = a_1 b_2 + a_2 b_3$ and $\varepsilon_{mn} = \varepsilon_m - \varepsilon_n$. $F_{\alpha\beta}(t) = \langle\psi_E(0)|U_{EI}^{(\beta)\dagger} U_{EI}^{(\alpha)}|\psi_E(0)\rangle$ is the decoherence factor [13,17], and in deriving the above equation we have used the obvious fact that $F_{\alpha 1} = F_{\alpha 2}$ and $F_{1\beta} = F_{2\beta}$ ($\alpha, \beta = 0, 1, 2, 3$). Moreover, in the absence of the self-Hamiltonian $\hat{H}_S$ (i.e., the two central spins have no interactions with each other), we have $b_1 = b_3 = 1$ and $b_2 = 0$, which yields $\mu_1 = a_1$ and $\mu_2 = a_2$.

Let us consider in detail the decoherence factor

$$F_{\alpha\beta}(t) = \langle\psi_E(0)|U_{EI}^{(\beta)\dagger} U_{EI}^{(\alpha)}|\psi_E(0)\rangle = \prod_{k>0} F_k, \tag{19}$$

which reflects the overlap between the two states of the spin environment obtained by evolving the initial state $|\psi_E(0)\rangle$ with two unitary operators $U_{EI}^{(\alpha)}$ and $U_{EI}^{(\beta)}$. From the time evolution operator $U_{EI}(t)$ expressed in Eq. (12) and the initial vacuum environment state, we obtain

$$\begin{aligned}|F_{\alpha\beta}(t)| = \prod_{k>0}\{&1 - \sin^2(\Omega_k^{(\alpha)}t)\sin^2(\Omega_k^{(\beta)}t)\sin^2(\theta_k^{(\alpha)} - \theta_k^{(\beta)}) - [\sin(\Omega_k^{(\alpha)}t) \\ &\times \cos(\Omega_k^{(\beta)}t)\sin(\theta_k^{(\alpha)}) - \cos(\Omega_k^{(\alpha)}t)\sin(\Omega_k^{(\beta)}t)\sin(\theta_k^{(\beta)})]^2\}^{1/2},\end{aligned} \tag{20}$$

where $\Omega_k^{(n)}$ and $\theta_k^{(n)}$ ($n = \alpha, \beta$) can be obtained by replacing $\hat{\Lambda}$ with $\Lambda_n$ in Eqs. (7) and (9), respectively. Here $\Lambda_{0,3} = \lambda \pm g$ and $\Lambda_{1,2} = \lambda$. Obviously, when $\gamma = 0$, $\sin(\theta_k^{(n)}) = 0$, as a result, one has $|F_{\alpha\beta}(t)| \equiv 1$, i.e., for the environment described by the *XX* model, the central spin system will be unaffected by its surrounding environment.

We now make a heuristic analysis of the features of the decoherence factors $F_{01}(t)$, $F_{03}(t)$ and $F_{13}(t)$. For this purpose, we introduce a cutoff frequency $K_c$ and define the partial product for $F_{\alpha\beta}(t)$ as [7,13]

$$|F_{\alpha\beta}(t)|_c = \prod_{k>0}^{K_c} F_k \geqslant |F_{\alpha\beta}(t)|, \tag{21}$$

from which the corresponding partial sum can be readily obtained as

$$S_{\alpha\beta}(t) = \ln|F_{\alpha\beta}(t)|_c = -\sum_{k>0}^{K_c} |\ln F_k|. \tag{22}$$

For the case that $N$ is large enough and $k$ is small relatively, one has $\epsilon_k^{(n)} \approx |1 - \Lambda_n|$ and $\Omega_k^{(n)} \approx 2(\epsilon_k^{(n)} - 2D x_k)$, consequently

$$\sin^2(\theta_k^{(\alpha)} - \theta_k^{(\beta)}) = \frac{4\pi^2 k^2 \gamma^2 (\Lambda_\alpha - \Lambda_\beta)^2}{N^2 (1-\Lambda_\alpha)^2 (1-\Lambda_\beta)^2}. \tag{23}$$

As a result, the approximation of $S(t)$ can be obtained as

$$S_{\alpha\beta}(t) \approx -\frac{2\pi^2 \gamma^2}{N^2 (1-\Lambda_\alpha)^2 (1-\Lambda_\beta)^2} \sum_{k>0}^{K_c} k^2 \{(\Lambda_\alpha - \Lambda_\beta)^2 \sin^2(\Omega_k^{(\alpha)} t) \sin^2(\Omega_k^{(\beta)} t) \\ + [\sin(\Omega_k^{(\alpha)} t) \cos(\Omega_k^{(\beta)} t)|1-\Lambda_\beta| - \cos(\Omega_k^{(\alpha)} t) \sin(\Omega_k^{(\beta)} t)|1-\Lambda_\alpha|]^2\}. \tag{24}$$

In the derivation of the above equation, we have used the equality $\ln(1-x) \approx -x$ for very small $x$.

For our two-qubit case, $\Lambda_{0,3} = \lambda \pm g$ and $\Lambda_{1,2} = \lambda$, thus when $\lambda$ is adjusted to the vicinity of the critical point $\lambda_c = 1$, and in the weak coupling regime $g \ll 1$ one has

$$|F_{01}(t)|_c = |F_{13}(t)|_c \approx e^{-\tau_1 t^4}, \ |F_{03}(t)|_c \approx e^{-\tau_2 t^4}, \tag{25}$$

for short time $t$, with $\tau_1$ and $\tau_2$ given by

$$\tau_1 = 8\gamma^2 \left[ E^{(2)}(K_c) g^2 - \frac{4 E^{(3)}(K_c)(\delta + g) g D}{\delta} \right], \\ \tau_2 = 32 E^{(2)}(K_c) \gamma^2 g^2 - 256 E^{(3)}(K_c) \gamma^2 g D, \tag{26}$$

where we have written $\delta = |\lambda - 1|$ and $E^{(n)}(K_c) = (2\pi/N)^n \Sigma_{k>0}^{K_c} k^n$ for concision. $E^{(n)}(K_c)$ are given by $E^{(2)}(K_c) = 4\pi^2 K_c(K_c+1)(2K_c+1)/(6N^2)$ and $E^{(3)}(K_c) = 8\pi^3 K_c^2 (K_c+1)^2 / (4N^3)$, respectively. In the derivation of $\tau_1$ and $\tau_2$, we have ignored the terms related to the sums of $k^4/N^4$, $k^5/N^5$ and $k^6/N^6$. From these approximation analysis one can found that when the magnetic field $\lambda$ is adjusted to the vicinity of the critical point $\lambda_c = 1$, the decoherence factors will exponentially decay with the fourth power of time. This is different from that of Ref. [17], which is certainly due to the different initial state of the environment we have chosen. Moreover, since $E^{(2)}(K_c) \gg E^{(3)}(K_c)$, introducing the DM interaction can only change the decay of the decoherence factors slightly in the weak coupling regime $g \ll 1$. For the special case $D = 0$ one has $\tau_2 = 4\tau_1 = 32 E^{(2)}(K_c) \gamma^2 g^2$, which indicates that in the absence of the DM interaction, $|F_{03}(t)|_c$ decays about four times as rapid as that of $|F_{01}(t)|_c = |F_{13}(t)|_c$.

In the strong coupling regime $g \gg 1$, from Eq. (24) one can derive the three partial sums directly as $S_{01}(t) = S_{13}(t) \approx -[2E^{(2)}(K_c) - 8E^{(3)}(K_c)D/\delta]\gamma^2 t^2$ and $S_{03}(t) \to 0$, from which the relevant decoherence factors can be obtained as $|F_{01}(t)|_c = |F_{13}(t)|_c = e^{-\tau_0 t^2}$ and $|F_{03}(t)|_c \approx 1$, where $\tau_0 = [2E^{(2)}(K_c) - 8E^{(3)}(K_c)D/\delta]\gamma^2$. Clearly, $|F_{01}(t)|_c = |F_{13}(t)|_c$ will decay exponentially with the second power of time, while $|F_{03}(t)|_c$ keeps a constant value of unity and no longer decays to zero exponentially. In fact, when $g \gg 1$, from the angle of Bogoliubov transformation expressed in Eq. (9) one can see that $\theta_k^{(0)} \approx \pi$ and $\theta_k^{(3)} \approx 0$, combination of this with Eq. (20) gives rise to $|F_{03}(t)|_c \approx 1$.

In the following, we check in detail the influence of the environment on the disentanglement dynamics of the central two-qubit system by resorting to numerical calculation. We first consider the initial state $|\psi_S\rangle = (|00\rangle + |11\rangle)/\sqrt{2}$, for which the concurrence [25] of the reduced density matrix that quantifies the degree of the pairwise entanglement between the two central qubits can be obtained explicitly as $C = |F_{03}|$. Obviously, the concurrence $C$ equals to the norm of the decoherence factor $F_{03}$, and is independent of the relative parameters of the self-Hamiltonian $\hat{H}_S$. This indicates that the entanglement of the two interacting central qubits must obey the same dynamical behaviors as that of the non-interacting ones for the initial state $|\psi_S\rangle$. Moreover, the purity [7] of $|\psi_S\rangle$ is given by $P = (1+|F_{03}|^2)/2 = (1+C^2)/2$, which is also determined by the norm of the decoherence factor $F_{03}$ only.

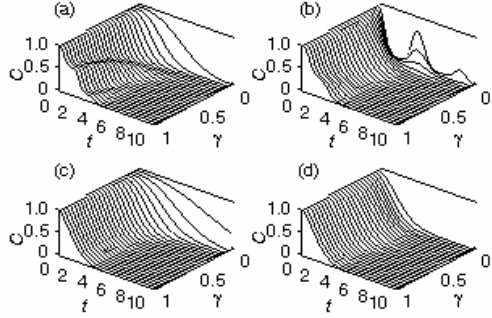

**Fig. 1.** Concurrence $C$ versus the anisotropic parameter $\gamma$ and time $t$ in the weak coupling regime $g = 0.02$ with the size of the environment given by $N = 1001$. (a) $D = 0$, $\lambda = 0.05$; (b) $D = 0.2$, $\lambda = 0.05$; (c) $D = 0$, $\lambda = 1$; (d) $D = 0.2$, $\lambda = 1$.

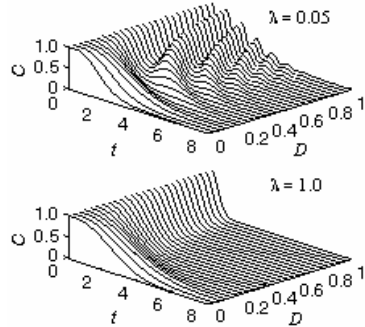

**Fig. 2.** Concurrence $C$ versus the DM interaction $D$ and time $t$ in the weak coupling regime $g = 0.02$ with the size of the environment $N = 1001$ and anisotropic parameter $\gamma = 0.2$.

To see the parameter-dependence of the disentanglement process in detail, in Fig. 1 we plot the concurrence versus the anisotropic parameter $\gamma$ of the spin environment and evolution time $t$ in the weak coupling regime $g = 0.02$ with different values of the DM interaction and magnetic field.

For the environment described by the *XX* model (i.e., $\gamma = 0$), one can find that the concurrence of the two central qubits remains in unity during its time evolution process. As a consequence, the purity of the two central qubits $P = (1 + C^2)/2$ also remains in unity, results in the disappearance of the coupling-induced decoherence. For the environment described by the general *XY* model (i.e., $\gamma \neq 0$), however, the concurrence decays with time, and this decaying behavior becomes more sharply when increasing the in-plane anisotropy of the environmental spin chain. In particular, in the vicinity of the critical point $\lambda_c = 1$, the concurrence decays monotonously as time evolves (this is consistent with the above heuristic analysis), while out of this region, it displays relatively complex behaviors (see also the top panel of Fig. 2).

By comparing the left two panels and the right two panels of Fig.1, one can also see that the decay of the concurrence may be enhanced by increasing the intensity of the DM interaction for spin-chain environment with small but nonzero in-plane anisotropy. Similar conclusions were also found by Cheng *et al.* [17] and Guo *et al.* [18] when studying decoherence problem with however different initial state of the environment. To show this phenomenon more clearly, in Fig. 2 we plot the concurrence versus the DM interaction $D$ and time $t$ in the weak coupling regime $g = 0.02$. Obviously, with relative weak magnetic field (e.g., $\lambda = 0.05$), the concurrence displays oscillating behaviors as time evolves when $D$ is larger than a certain value $D_c$, however, when $\lambda$ takes the critical value $\lambda_c = 1$, the concurrence always decay monotonously as time evolves, irrespective of the intensity of the DM interaction.

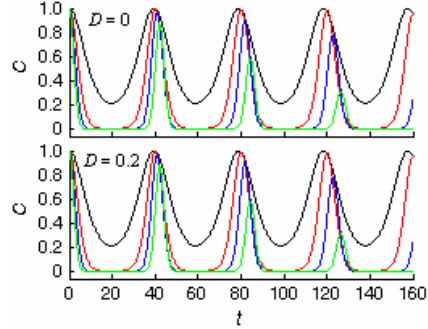

**Fig. 3.** (Color online) Concurrence $C$ versus time $t$ in the weak coupling regime $g = 0.02$ with $N = 1001$ and $\lambda = 2$. For every plot the curves from top to bottom correspond to $\gamma = 0.2, 0.4, 0.6$ and $0.8$.

In Fig. 3 we show disentanglement dynamics of the two-qubit state $|\psi_S\rangle$ with magnetic field intensity $\lambda$ stronger than $\lambda_c = 1$ with different values of the anisotropic parameter $\gamma$ and DM interaction $D$, from which one can find that the concurrence $C$ oscillates as the time evolves, with the phenomenon of entanglement sudden death (ESD) [2] and birth being observed when $\gamma$ becomes larger than a critical value. This is different from that for small $\lambda$, in which no birth of entanglement can be found (see Fig. 1). The plot also indicates that the effects of increasing the intensity of the DM interaction on the decay of the concurrence is not remarkable for strong magnetic field.

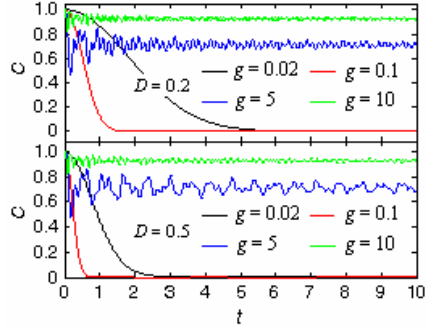

**Fig. 4.** (Color online) Concurrence $C$ versus time $t$ at the critical point $\lambda_c = 1$ with different coupling strengths $g$ and DM interaction $D$. The other parameters for the plot are $N = 1001$ and $\gamma = 0.2$.

Fig. 4 shows the effects of increasing the coupling strength $g$ between the central spin system and its environmental spin chain on disentanglement dynamics of the two-qubit state. It can be found that for fixed intensity of the DM interaction, if one properly enlarges the coupling strength (e.g., $g = 0.1$), the concurrence evolves from unity to zero in a more short time compared with that of $g = 0.02$, which implies that the disentanglement of the two central qubits is best enhanced. However, if one continues enlarging the coupling strength (e.g., $g = 5$), the concurrence begins to oscillating rapidly around a steady value which increases as the anisotropy of the environmental spin chain decreases. For the case of very strong coupling strength (e.g., $g = 10$), the concurrence behaves as a weak oscillation near the maximum value $C_{\max} = 1$, from which it is reasonable to conjecture that up to the strong coupling limit case, the concurrence will keep an invariant value of unity during its time evolution process. Moreover, by comparing the two panels of Fig. 4, one can also found that the decay of the concurrence may be enhanced by increasing the intensity of the DM interaction for proper weak coupling strengths $g$ (e.g., $g = 0.02$ and $g = 0.1$).

Next we turn to investigate disentanglement process of another initial maximally entangled two-qubit state $|\psi_S\rangle = (|00\rangle + |01\rangle - |10\rangle + |11\rangle)/2$, and restrict our attention to the case that the parameters of the central spin system given by $\Delta = 1$ and $h_1 = h_2 = \lambda$ (i.e., the two central qubits correlated via the Heisenberg *XXX* interaction and are subjected to an uniform external magnetic field with strength the same as that of the environmental spin chain), which yields $\theta = \pi/2$, $\varepsilon_{0,3} = \Delta/4 \pm \lambda$, $\varepsilon_{1,2} = -\Delta/4 \pm 1/2$, $b_1 = b_3 = (e^{-i\varepsilon_1 t} + e^{-i\varepsilon_2 t})/2$, $b_2 = (e^{-i\varepsilon_1 t} - e^{-i\varepsilon_2 t})/2$ and $\mu_1 = -\mu_2 = e^{-i\varepsilon_2 t}/2$. Thus the reduced density matrix expressed in Eq. (18) simplifies to

$$\rho_S(t) = \frac{1}{4}\begin{pmatrix} 1 & e^{i\varepsilon_{20}t}F_{01} & -e^{i\varepsilon_{20}t}F_{01} & e^{i\varepsilon_{30}t}F_{03} \\ e^{i\varepsilon_{02}t}F_{01}^* & 1 & -1 & e^{i\varepsilon_{32}t}F_{13} \\ -e^{i\varepsilon_{02}t}F_{01}^* & -1 & 1 & -e^{i\varepsilon_{32}t}F_{13} \\ e^{i\varepsilon_{03}t}F_{03}^* & e^{i\varepsilon_{23}t}F_{13}^* & -e^{i\varepsilon_{23}t}F_{13}^* & 1 \end{pmatrix}, \qquad (27)$$

where we have used the notation $\varepsilon_{mn} = \varepsilon_m - \varepsilon_n$ for concision of the presentation. From this equation one can calculate the concurrence of this initial state. Moreover, the purity [7] of this state can be obtained analytically as $P = (3 + |F_{03}|^2 + 2|F_{01}|^2 + 2|F_{13}|^2)/8$. Different from those of the initial two-qubit pure state $|\psi_S\rangle = (|00\rangle + |11\rangle)/\sqrt{2}$, here both the concurrence and the purity are determined by the norm of the three decoherence factors $F_{01}$, $F_{03}$ and $F_{13}$, particularly, they are both dependent on the corresponding parameters of the central spin system.

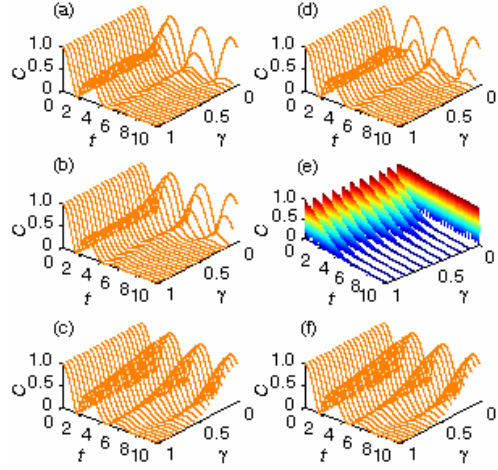

**Fig. 5.** (Color online) Concurrence $C$ versus the anisotropic parameter $\gamma$ and time $t$ in the weak coupling regime $g = 0.02$ with the environment size $N = 1001$. (a) $D = 0$, $\lambda = 0.05$; (b) $D = 0$, $\lambda = 1$; (c) $D = 0$, $\lambda = 2$; (d) $D = 0.2$, $\lambda = 0.05$; (e) $D = 0.2$, $\lambda = 1$; (f) $D = 0.2$, $\lambda = 2$.

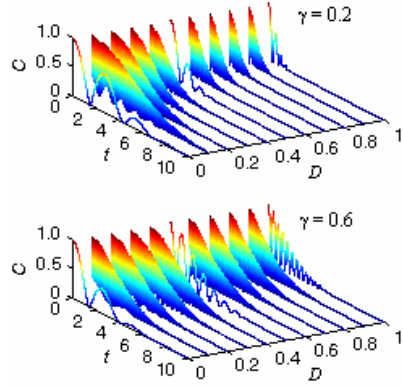

**Fig. 6.** (Color online) Concurrence $C$ versus the DM interaction $D$ and evolution time $t$ in the weak coupling regime $g = 0.02$ with $N = 1001$ and $\lambda_c = 1$.

In the weak coupling regime $g \ll 1$, one can see from Fig. 5 that when $\gamma = 0$, the concurrence evolves periodically between the maximal value 1 and the minimal value 0 with the period $\tau = \pi$ apart from the special case $(\lambda_c = 1, D \neq 0)$, for which the concurrence displays rapid oscillation behaviors in the whole time region. This behavior can be interpreted from the analytical results expressed in Eqs. (25) and (26), which yield $|F_{03}|_c = |F_{01}|_c = |F_{13}|_c = 1$ and the purity of the two central spins $P = (3 + |F_{03}|^2 + 2|F_{01}|^2 + 2|F_{13}|^2)/8 = 1$, thus the coupling-induced decoherence disappears for the environment described by the *XX* model. From Fig. 5 one can also see that if $\lambda \neq 1$, the influence of the DM interaction on the decay of the concurrence is not remarkable. At the critical point $\lambda_c = 1$, however, increasing the DM interaction has significant influence on the decay of the concurrence. For the anisotropic environment, when $D = 0$, the concurrence behaves as damped oscillations as the time evolves and the environment disentangling the two central spins completely in a finite time just as the "sudden death" of entanglement discovered previously by Yu and Eberly [2], while for the case of $D \neq 0$, the concurrence oscillates rapidly as the time evolves and the environment also disentangling the two central spins completely in a finite time. To reveal the effects of increasing the intensity of the DM interaction on the decay of the concurrence more clearly, we show in Fig. 6 the concurrence versus $D$ and $t$ at the critical point $\lambda_c = 1$. It can be found that for the special cases $D = 0$, $0.5$ and $1$ the concurrence vanishes after several simple damped oscillations, while for other values of $D$, it oscillates rapidly as time evolves and decays off to zero finally.

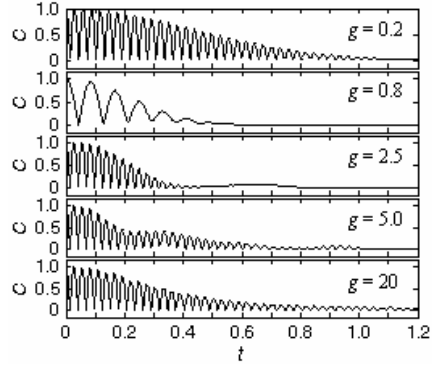

**Fig. 7.** Concurrence $C$ versus time $t$ at the critical point $\lambda_c = 1$ with different coupling strengths $g$. The other parameters for the plot are $N = 1001$, $\gamma = 0.2$ and $D = 0.2$.

In Fig. 7 we show effects of different coupling strengths $g$ on decay of the concurrence in the quantum critical region $\lambda_c = 1$. It can be found that the decay of the concurrence may be enhanced by smoothly tuning the value of $g$ (e.g., $g = 0.2$) a little larger than $g = 0.02$ [displayed in Fig. 5(e) and the top panel of Fig. 6], and this decaying behavior becomes even more remarkable with relative larger coupling strengths such as $g = 0.8$ and $2.5$. However, if one further increases the coupling strength $g$ (e.g., $g = 5$), the decay of the concurrence will be delayed slightly. To the strong coupling limit case, the decay of the concurrence show behaviors almost the same as that of $g = 20$, which are featured by damped oscillations as time evolves and decays off to zero finally. This is dramatically different from that of the initial state $|\psi_s\rangle = (|00\rangle + |11\rangle)/\sqrt{2}$, for which the concurrence will stay at $C_{\max} = 1$ without changing with time to the strong coupling limit.

## 4. Disentanglement dynamics of two qutrits

The effects of the DM interaction on the disentanglement dynamics of two central qutrits have been discussed recently [17,18], with however, the two qutrits have no intraspin interactions with each other. Here we will reconsider this problem by extending it to the correlated ones. The Hamiltonian of this central two qutrit system has the same form as that expressed in Eq. (2), with however, $s_1$ and $s_2$ representing the spin-1 operators. The eigenvalues of this Hamiltonian can be obtained as

$$\varepsilon_{0,1} = h \pm 1, \ \varepsilon_{2,3} = -h \pm 1, \ \varepsilon_{4,5} = \Delta \pm 2h, \ \varepsilon_6 = -\Delta, \tag{28}$$
$$\varepsilon_7 = \sqrt{2}\tan(\theta/2), \ \varepsilon_8 = -\sqrt{2}\cot(\theta/2),$$

with the corresponding eigenstates given by

$$|\varphi_{0,1}\rangle = (|01\rangle \pm |10\rangle)/\sqrt{2}, \ |\varphi_{2,3}\rangle = (|12\rangle \pm |21\rangle)/\sqrt{2},$$
$$|\varphi_4\rangle = |00\rangle, \ |\varphi_5\rangle = |22\rangle, \ |\varphi_6\rangle = (|02\rangle - |20\rangle)/\sqrt{2},$$
$$|\varphi_7\rangle = \sin(\theta/2)[|02\rangle + \sqrt{2}\cot(\theta/2)|11\rangle + |20\rangle]/\sqrt{2}, \tag{29}$$
$$|\varphi_8\rangle = \cos(\theta/2)[|02\rangle - \sqrt{2}\tan(\theta/2)|11\rangle + |20\rangle]/\sqrt{2},$$

where $\theta = \sin^{-1}(2\sqrt{2}/\sqrt{8+\Delta^2})$, with $|0\rangle$, $|1\rangle$ and $|2\rangle$ representing the spin-1 state with magnetic quantum number 1, 0 and $-1$, respectively, and we have chosen $h_1 = h_2 = h$ (i.e., the two central qutrits are subjected to an uniform external magnetic field) in deriving the above two equations.

In this section, we concentrate on the initial states of the two central qutrits and its surrounding environment as $|\psi_S\rangle = a_0|00\rangle + a_1|11\rangle + a_2|22\rangle$ and $|\psi_E(0)\rangle = |0\rangle_{k=0} \otimes_{k>0} |0\rangle_k |0\rangle_{-k}$, then by applying the time evolution operator $U(t)$, one can obtain the state vector at time $t$ as

$$|\Psi(t)\rangle = a_0 e^{-i\varepsilon_4 t}|00\rangle \otimes U_{EI}^{(0)}|\psi_E(0)\rangle + a_2 e^{-i\varepsilon_5 t}|22\rangle \otimes U_{EI}^{(2)}|\psi_E(0)\rangle$$
$$+ a_1\left[\tfrac{\sqrt{2}\sin\theta}{4}(e^{-i\varepsilon_7 t} - e^{-i\varepsilon_8 t})(|02\rangle + |20\rangle) + (\cos^2\tfrac{\theta}{2}e^{-i\varepsilon_7 t}\right.$$
$$\left.+ \sin^2\tfrac{\theta}{2}e^{-i\varepsilon_8 t})|11\rangle\right] \otimes U_{EI}^{(1)}|\psi_E(0)\rangle. \quad (30)$$

The unitary operators $U_{EI}^{(i)}$ ($i = 0, 1, 2$) can be obtained from the unitary operator $U_{EI}(t)$ by replacing the operator $\hat{\Lambda}$ with $\Lambda_{0,2} = \lambda \pm 2g$ and $\Lambda_1 = \lambda$. Tracing out the degrees of the spin environment, we obtained the reduced density matrix of the two central qutrits in the basis $\{|00\rangle, |11\rangle, |22\rangle, |02\rangle, |20\rangle, |01\rangle, |10\rangle, |12\rangle, |21\rangle\}$ as

$$\rho_S(t) = \begin{pmatrix} |a_0|^2 & a_0 a_1^* \alpha_2 F_{01} & a_0 a_2^* F_{02} e^{i\varepsilon_{54} t} & a_0 a_1^* \alpha_1 F_{01} & a_0 a_1^* \alpha_1 F_{01} \\ a_0^* a_1 \alpha_2^* F_{01}^* & |a_1|^2 \mu & a_1 a_2^* \beta_2^* F_{12} & |a_1|^2 \eta_2^* & |a_1|^2 \eta_2^* \\ a_0^* a_2 F_{02}^* e^{i\varepsilon_{45} t} & a_1^* a_2 \beta_2 F_{12}^* & |a_2|^2 & a_1^* a_2 \beta_1 F_{12}^* & a_1^* a_2 \beta_1 F_{12}^* \\ a_0^* a_1 \alpha_1^* F_{01}^* & |a_1|^2 \eta_2 & a_1 a_2^* \beta_1^* F_{12} & |a_1|^2 \eta_1 & |a_1|^2 \eta_1 \\ a_0^* a_1 \alpha_1^* F_{01}^* & |a_1|^2 \eta_2 & a_1 a_2^* \beta_1^* F_{12} & |a_1|^2 \eta_1^* & |a_1|^2 \eta_1 \end{pmatrix} \oplus Z_{4\times 4}, \quad (31)$$

where we still used the notation $\varepsilon_{mn} = \varepsilon_m - \varepsilon_n$ for concision of the presentation, and $Z_{4\times 4}$ denotes the $4 \times 4$ zero matrix. The other parameters in $\rho_S(t)$ are given by

$$\alpha_1 = \tfrac{\sqrt{2}\sin\theta}{4}(e^{-i\varepsilon_{47} t} - e^{-i\varepsilon_{48} t}), \quad \alpha_2 = \cos^2\tfrac{\theta}{2} e^{-i\varepsilon_{47} t} + \sin^2\tfrac{\theta}{2} e^{-i\varepsilon_{48} t},$$
$$\beta_1 = \tfrac{\sqrt{2}\sin\theta}{4}(e^{-i\varepsilon_{57} t} - e^{-i\varepsilon_{58} t}), \quad \beta_2 = \cos^2\tfrac{\theta}{2} e^{-i\varepsilon_{57} t} + \sin^2\tfrac{\theta}{2} e^{-i\varepsilon_{58} t}, \quad (32)$$
$$\eta_1 = \tfrac{\sin^2\theta}{4}[1 - \cos(\varepsilon_{78} t)], \quad \eta_2 = \tfrac{\sqrt{2}\sin\theta}{4}(\cos\theta + \sin^2\tfrac{\theta}{2} e^{-i\varepsilon_{78} t} - \cos^2\tfrac{\theta}{2} e^{i\varepsilon_{78} t}),$$
$$\mu = \sin^4\tfrac{\theta}{2} + \cos^4\tfrac{\theta}{2} + 2\sin^2\tfrac{\theta}{2}\cos^2\tfrac{\theta}{2}\cos(\varepsilon_{78} t).$$

Here $F_{\alpha\beta}(t) = \langle \psi_E(0)|U_{EI}^{(\beta)\dagger}U_{EI}^{(\alpha)}|\psi_E(0)\rangle$ represents the decoherence factor. Since $\Lambda_{0,3} = \lambda \pm g$, $\Lambda_{1,2} = \lambda$ for the two-qubit system, and $\Lambda_{0,2} = \lambda \pm 2g$, $\Lambda_1 = \lambda$ for the two-qutrit system, the three decoherence factors $F_{01}(t)$, $F_{02}(t)$ and $F_{12}(t)$ presented in Eq. (32) should obey similar dynamical behaviors as those of $F_{01}(t)$, $F_{03}(t)$ and $F_{13}(t)$ for the two qubits.

In this paper, we adopt the concept of negativity [26] to measure the entanglement between the two central qutrits. For this purpose, we calculate the partial transpose of $\rho_S(t)$ with respect to the second subsystem, which yields $\rho_S^{T_2}(t) = |a_1|^2 \mu \oplus B_1 \oplus B_2$, with

$$B_1 = \begin{pmatrix} |a_0|^2 & |a_1|^2 \eta_1 & a_0^* a_1 \alpha_1^* F_{01}^* & a_0 a_1^* \alpha_1 F_{01} \\ |a_1|^2 \eta_1^* & |a_2|^2 & a_1^* a_2 \beta_1 F_{12}^* & a_1 a_2^* \beta_1^* F_{12} \\ a_0 a_1^* \alpha_1 F_{01} & a_1 a_2^* \beta_1^* F_{12} & |a_1|^2 \eta_1 & a_0 a_2^* e^{-i\varepsilon_{45} t} F_{02} \\ a_0^* a_1 \alpha_1^* F_{01}^* & a_1^* a_2 \beta_1 F_{12}^* & a_0^* a_2 e^{i\varepsilon_{45} t} F_{02}^* & |a_1|^2 \eta_1 \end{pmatrix},$$

$$B_2 = \begin{pmatrix} 0 & a_0 a_1^* \alpha_2 F_{01} & |a_1|^2 \eta_2 & 0 \\ a_0^* a_1 \alpha_2^* F_{01}^* & 0 & 0 & |a_1|^2 \eta_2^* \\ |a_1|^2 \eta_2^* & 0 & 0 & a_1 a_2^* \beta_2^* F_{12} \\ 0 & |a_1|^2 \eta_2 & a_1^* a_2 \beta_2 F_{12}^* & 0 \end{pmatrix}.$$

(33)

When there is no self-Hamiltonian (i.e., $\hat{H}_S = 0$), we have $\alpha_1 = \beta_1 = \eta_1 = \eta_2 = 0$ and $\alpha_2 = \beta_2 = \mu = 1$. Clearly, for this special case, $\rho_S^{T_2}(t)$ reduces to that expressed in Eq. (33) of Ref. [13].

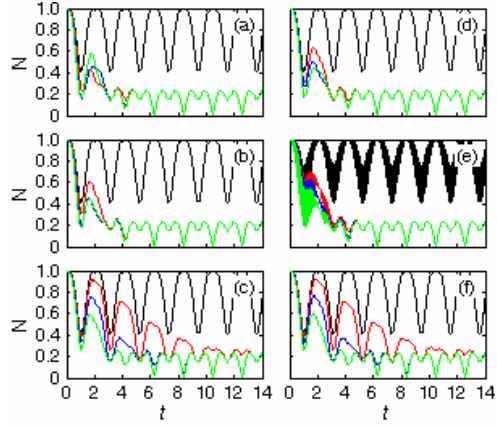

**Fig. 8.** (Color online) Negativity $\mathcal{N}$ versus time $t$ in the weak coupling regime $g = 0.02$ with the environment size $N = 601$. For every plot, the black, red, blue and green curves correspond to the case $\gamma = 0, 0.3, 0.6$ and $1$, respectively. (a) $D = 0$, $\lambda = 0.05$; (b) $D = 0$, $\lambda = 1$; (c) $D = 0$, $\lambda = 2$; (d) $D = 0.2$, $\lambda = 0.05$; (e) $D = 0.2$, $\lambda = 1$; (f) $D = 0.2$, $\lambda = 2$.

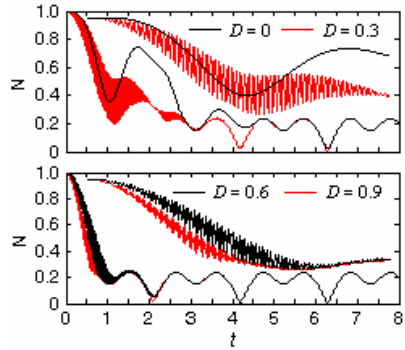

**Fig. 9.** (Color online) Negativity $\mathcal{N}$ versus time $t$ at the critical point $\lambda_c = 1$ with different DM interactions. The other parameters for the plot are $N = 601$, $\gamma = 0.2$ and $g = 0.02$. The insets in the top and bottom panels show dynamics of the negativity during the time intervals $t \in [0, 2]$ and $t \in [0, 1.5]$.

Eqs. (31), (32) and (33) enable us to calculate the negativity [26] of the two central qutrits straightforwardly. The numerical results for different system-environment parameters are shown in Figs. 8, 9 and 10, where we have chosen $h = \lambda$, $\Delta = 1$ and $a_0 = a_1 = a_2 = 1/\sqrt{3}$ in all these plots.

- 13 -

Fig. 8 shows the negativity versus time $t$ in the weak coupling regime $g = 0.02$ with different strengths of the DM interaction and magnetic field. It can be observed that the negativity displays completely different behaviors compared with that of the uncorrelated two central qutrits [13,17]. In particular, the enhancement of the disentanglement of the two qutrits occurred at the critical point of the environment [13] is broken by introducing self-Hamiltonian of the central system. For the isotropic spin-chain environment (i.e., $\gamma = 0$), the negativity oscillates periodically between the minimal value $11/27$ [occurs at $t_c = (1+2k)\pi/3, k \in Z$] and the maximal value 1 [occurs at $t_c = 2k\pi/3, k \in Z$] with the period given by $\tau = 2\pi/3$, and displays completely the same dynamical behaviors for different values of $\lambda$ and $D$ apart from the special case $(\lambda_c = 1, D \neq 0)$, for which the negativity oscillates periodically accompanied by rapid sub-oscillations in the whole time region. For the anisotropic spin-chain environment (i.e., $\gamma \neq 0$), the decay of the negativity is affected remarkably by varying the in-plane anisotropy of the environment only in the short-time region, while in the long-time region, the effects of varying the in-plane anisotropy on decay of the negativity is neglectable. Moreover, from Fig. 8 one can also observe that when the magnetic field is adjusted to the critical point $\lambda_c = 1$, the decay of the negativity is affected significantly by introducing the DM interaction. When $D \neq 0$, the negativity oscillates rapidly as the time evolves in the short-time region if $\gamma \neq 0$. Same behaviors are demonstrated more clearly in Fig. 9, from which one can see that the regions of validity of rapid oscillations of the negativity is shortened slightly with increasing value of the DM interaction, with the amplitude of rapid sub-oscillations also decreases as $D$ increases. Further study reveals that only when the intensity of the transverse magnetic field $\lambda \in (\lambda_c - 2g, \lambda_c + 2g]$ can the negativity displays rapid oscillation behaviors in the short-time region. To the long time limit case, as shown in Fig. 9, the influence of the DM interaction on the decay of the negativity is, however, not remarkable.

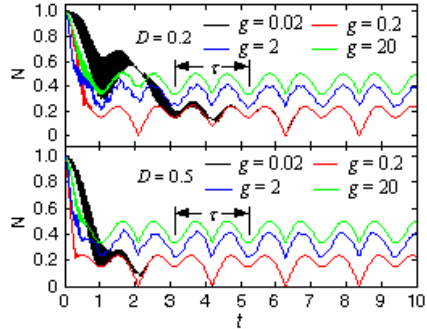

**Fig. 10.** (Color online) Negativity $\mathcal{N}$ versus time $t$ at the critical point $\lambda_c = 1$ with different coupling strengths $g$ and DM interaction $D$. The other parameters for the plot are $N = 601$ and $\gamma = 0.2$.

In Fig. 10, we numerically investigate influence of increasing the coupling strength $g$ on decay of the negativity at the critical point $\lambda_c = 1$. For fixed anisotropic parameter and DM interaction, if one increases the coupling strength (e.g., $g = 0.2$) moderately, both the region in which the negativity displays rapid sub-oscillation behaviors and the amplitude of rapid sub-oscillations are affected remarkably in the short-time region, while in the long-time region, the negativity displays almost the same dynamical behaviors as that of $g = 0.02$. If one further increases the coupling strength (e.g., $g = 2$), however, the value of the negativity will be increased, and to the strong coupling limit case, the negativity will decay from the initial maximal value 1 to a small value of about $1/3$ in a short time and then begin to oscillating periodically between the minimal value $1/3$ and the maximal value $1/2$ in the subsequent evolution, with the period given by $\tau = 2\pi/3$, which is independent of the strength of the DM interaction.

## 5. Conclusion

In summary, we have investigated the disentanglement process of two qubits and two qutrits uniformly coupled to a general *XY* spin chain (serves as a many-body surrounding environment) with the DM interaction. Different from the previous works [11–15,17,18] in which the authors concentrated on uncorrelated central spin systems, in this paper we discussed disentanglement of two central spins coupled via the *XXZ* interaction, our results revealed many novel phenomena of disentanglement dynamics.

For the initial two-qubit state, our heuristic analysis revealed that in the weak coupling regime $g \ll 1$, the norm of the decoherence factors $F_{\alpha\beta}(t)$ present an exponentially decay with the fourth power of time in the vicinity of the critical point $\lambda_c = 1$, and introducing the DM interaction can only change the decay slightly. In the strong coupling regime $g \gg 1$, however, $|F_{01}(t)| = |F_{13}(t)|$ decay exponentially with the second power of time, while $|F_{03}(t)|$ keeps a constant value of unity. As concrete examples, we examined disentanglement of two initial maximally entangled states. For $|\psi_S\rangle = (|00\rangle + |11\rangle)/\sqrt{2}$, it was found that in the weak coupling regime $g \ll 1$, the decay of the concurrence is enhanced by increasing the DM interaction when the two qubits are exposed to spin environment with small in-plane anisotropy. For $|\psi_S\rangle = (|00\rangle + |01\rangle - |10\rangle + |11\rangle)/2$, only when $\lambda_c = 1$ can the decay of the concurrence be affected significantly by introducing the DM interaction. Moreover, the concurrence displays relative simple behaviors for $D = 0$, 0.5 and 1 compared with other cases of $D$ in the weak coupling regime $g \ll 1$.

We also studied disentanglement dynamics of two correlated qutrits coupled to the spin-chain environment. When the system evolves from the initial state $|\psi_S\rangle = (|00\rangle + |11\rangle + |22\rangle)/\sqrt{3}$, we found that in the vicinity of the critical point $\lambda_c = 1$, the decay of the negativity is affected significantly by increasing the intensity of the DM interaction for either the weak or strong coupling case in the short-time region, while up to the long time limit, the influence of the DM interaction on decay of the negativity is not remarkable.

The role of the DM interaction played on the decay of the entanglement has also been discussed previously by Cheng *et al.* [17] and Guo *et al.* [18], however, our present study differs in two aspects from those of [17,18]. First, we assumed different initial state of the spin-chain environment, and second, we emphasized on disentanglement process of the central system with spins that have interactions with each other (i.e., $\hat{H}_S \neq 0$). Our study revealed many unexplored phenomena that are nontrivially different from those of the uncorrelated ones, in particular, the enhanced decay of the entanglement occurred at the critical point of the surrounding environment [13,14] may be broken by introducing self-Hamiltonian of the central system. We hope these results will assist in gaining further understanding of the mechanism of the decoherence induced by the spin-chain environment and its relation to various QIP tasks.

## Acknowledgments

This work was supported by the Natural Science Foundation of Shaanxi Province under Grant Nos. 2010JM1011 and 2009JQ8006, the Specialized Research Program of Education Department of Shaanxi Provincial Government under Grant No. 08JK434.